\documentstyle[11pt,newpasp,twoside]{article}
\begin{document}

\title{Enhanced interactions, mergers and starbursts in dense
cluster environments}
\author{C. Moss}
\affil{Astrophysics Research Institute, Liverpool John Moores University,
Twelve Quays House, Egerton Wharf, Birkenhead CH41 1LD, U.K.}

\begin{abstract}

Optical and X-ray studies have established the prevalence of
significant substructure in clusters of galaxies, indicating that
clusters are young systems, and that recent major mergers have
occurred in many clusters.  Numerical simulations show that
sub-cluster merging will result in significant tidal forces on disk
galaxies in accreting sub-groups, which is likely to lead to the
transformation of spirals to S0s in clusters.  Simulations predict
simultaneous moderate starbursts in gas-rich disk galaxies
in clusters which show on-going or recent merger activity, as well as
enhanced galaxy--galaxy interactions and mergers.
Observational data from studies of spectral indicators, both of starburst
activity in late-type galaxies in nearby clusters, and of poststarburst
activity in early-type galaxies in nearby and intermediate redshift
clusters, support this scenario.

\end{abstract}

\section{Introduction: sub-cluster merging and cluster
disk galaxy evolution}

Both optical studies of sky-projected cluster galaxy distributions
(e.g. Baier 1979; Geller \& Beers 1982) and studies of cluster X-ray
morphologies (e.g. Mohr, Fabricant \& Geller 1993; Buote \& Tsai 1996)
reveal the presence of significant substructure in clusters of
galaxies. Subgroup velocity statistics (e.g. Dressler \& Shectman
1988; Bird 1994; Girardi et al. 1997; Solanes, Salvador-Sol\'{e} \&
Gonz\'{a}lez-Casado 1999) confirm these results, and are in substantial
agreement in showing that some 30--40\% of clusters have statistically
significant substructure.

Numerical simulations also show that subclustering is a typical
property of cluster formation in hierarchical theories of structure
formation.  Such subclustering is expected for a broad
range of cosmologies, and its predicted extent is in
agreement with that observed (e.g. Knebe \& M\"{u}ller 2000). Its
prevalance indicates that clusters are young objects,
since such substructure is expected to be smoothed out over a few
cluster crossing times.  Moreover the existing substructure indicates
that recent major mergers have occurred in many clusters.

The evolution of cluster substructure is potentially of great
importance for its influence on the star formation histories and
morphological evolution of cluster galaxies (e.g Tully \& Shaya 1984;
Lavery \& Henry 1988; Gnedin 1999; Bekki 1999).  However, until very
recently this potential influence has largely been neglected
(cf. Bekki 1999). Of particular interest, is its relevance to the
evolution of disk galaxies in clusters.

As is well known, Butcher \& Oemler (1978) found that the cores of
intermediate redshift clusters contain a higher fraction of blue,
star-forming galaxies than similar environments at the present. These
galaxies have been shown to be normal spiral or irregular galaxies, a
fraction of which are obviously interacting or disturbed
(e.g. Dressler et al. 1994; Oemler, Dressler \& Butcher 1997; Smail et
al. 1997).  In rich clusters they constitute up to 50\% of the
population, but by the present epoch have been depleted by a factor of
two and replaced by a corresponding increase in S0s (e.g. Dressler et
al. 1997).  Wang \& Ulmer (1997) have found a correlation between the
fraction of blue galaxies in intermediate redshift clusters and the
cluster X-ray ellipticities.  They interpret the latter as an age
sequence, and conclude that the younger clusters have a higher
fraction of blue, star-forming galaxies. As the authors note, this
interpretation is in accord with a hierarchical clustering scenario,
in which a typical distant cluster is assembled from smaller
units which tend to contain more gas-rich, late type galaxies.  The
relative depletion of such star-forming galaxies in older, more
relaxed clusters, suggests that the process of sub-cluster merging
plays a crucial role in the transformation of an initial population of
cluster spirals to a subsequent S0 population.

What effect is the sub-cluster merging process likely to have on a
population of gas-rich, star-forming disk galaxies? Bekki (1999) using
numerical simulation of a merger between a small group of galaxies and
a cluster, shows that the time-dependent tidal gravitational field of
the merger gives strong nonaxisymmetric perturbations to the disk
galaxies in the group, inducing simultaneous moderate secondary
starbursts in the central regions of the galaxies.  

Similarly, Gnedin (1999) using self-consistent cluster simulations,
demonstrates that for substantial infall of additional material into a
cluster, the time-varying gravitational potential causes a sequence of
tidal shocks on individual galaxies over a wide area of the cluster.
These shocks enhance galaxy--galaxy interactions as well as amplifying
galaxy merger rates.  Again the effect is also likely to induce
simultaneous starbursts in gas-rich infalling disk galaxies over a
wide area of the cluster. Moreover, the tidal shocks are shown to
typically thicken the disks of large spirals by a factor of two,
making it unlikely that spiral structure and gaseous shocks in the
inner regions of the galaxies will form. A significant amount of
the dark matter halo is stripped, possibly removing the resevoir of
fresh gas which maintains the star formation activity (cf. Larson,
Tinsley \& Caldwell 1980).  While gas in the outer regions of the disk
can be stripped by the ram pressure of the ambient medium,
interpenetrating collisions with neighbouring galaxies could remove
all their gas.  The effect of all these tidal transformations is the
conversion of spirals to S0s.

\section{Observational evidence for tidally-induced starbursts in cluster
galaxies}

The theoretical considerations discussed above suggest that (moderate)
tidally-induced starbursts should be widespread in cluster gas-rich disk
galaxies in clusters which show evidence of on-going
or recent merger activity.  Together with enhanced starburst activity,
enhanced galaxy--galaxy interactions and mergers are also expected.
In fact, there is considerable observational evidence to support this
scenario, both from studies of early-type and late-type galaxies in
nearby clusters, and from evidence of poststarburst galaxies in 
intermediate redshift clusters. This evidence is now briefly reviewed.  

Caldwell et al. (1993) have surveyed early-type galaxies in the Coma
cluster and found an unusually high number with spectra that reflect
recent enhanced star formation activity in a substructure of the
cluster. As noted by Bekki (1999), this may readily be explained as
due principally to tidal gravitational effects of a group-cluster
merger with rather large relative velocity.  Caldwell \& Rose (1997)
also found a larger fraction of poststarburst galaxies in nearby
clusters with obvious double structure.  In a similar fashion, Bekki
notes that these results can also be explained by the tidal effects of
cluster mergers.

Moss \& Whittle (2000) have undertaken an H$\alpha$ survey of a
magnitude-limited complete sample of 320 galaxies of types Sa and
later within 1.5$h^{-1}$ Mpc of the centres of 8 nearby Abell
clusters. Some 116 of the sample were detected in emission.  A subset
of the emission-line galaxies are found to have tidally-induced
circumnuclear starburst emission, as evidenced from the particular
H$\alpha$ morphology, and its strong correlation with a disturbed
galaxy morphology. Starburst emission is found to be most prevalent in
the richest clusters in the survey.  The percentage of the total
sample of late-type galaxies with starburst emission is $\sim$ 40\%
for the two richest clusters, Coma and Abell 1367. In contrast, the
corresponding percentage for field galaxies in the sample is only a
few percent.  Moreover X-ray morphologies and temperature structures
for Coma and Abell 1367 indicate that both these clusters are
recent post-merger systems (cf. Donnelly et al. 1998; Honda et
al. 1996).  Again, a plausible explanation for the enhanced starburst
activity seen in the clusters, is tidal effects of subcluster merging.

Moss \& Whittle also identify a class of late-type galaxies in their
sample which are likely to be on-going mergers, in which the companion
is indistinguishable from its merger partner.  These on-going mergers
are most prevalent in the Coma cluster. They are likely to represent a
later stage of close double, interacting systems with tidally induced
emission which comprise a significant fraction ($\sim$ 14\%) of the
galaxies they detect with starburst emission.

For intermediate redshift clusters, there is also evidence of an
enhancement of starburst emission.  Galaxies with spectra showing
strong Balmer lines in absorption, indicative of a poststarburst
phase, are an order of magnitude more frequent in the cluster
environment as compared to the high redshift field (e.g. Dressler et
al. 1999; Poggianti et al. 1999).  A surprising fraction of the
star-forming galaxies at intermediate redshift show signs of
morphological disturbance, of which many are suggestive of merging
systems, virtually always involving a disk galaxy (cf. Ellis 1999).
This suggests a tidal origin for the starburst phase,
although no clear difference in the incidence of disturbed morphology
between the high redshift field and cluster environments has been
established (cf. Dressler et al. 1999).

\section{Conclusion}

Strong theoretical and observational data suggest that sub-cluster
merging plays a crucial role in the transformation of spirals to S0s
in the process of cluster formation, due to the strong tidal forces on
the disk galaxy members of an accreting sub-group. Theoretical studies
predict simultaneous moderate starbursts in disk galaxies of the
accreting sub-group due to tidal forces.  This prediction is supported
by observational data for both nearby and intermediate redshift
clusters.  Tidal shocks due to the time-varying potential associated
with sub-cluster merging are expected to effect morphological
transformation of star-forming disk galaxies to S0s.  Thus these can
explain the transformation of the majority of the spiral population in
younger, less relaxed clusters to a population of S0s in older, more
relaxed clusters as an integral part of the process of cluster
formation.

 \end{document}